# Lewis Acidity and Basicity Diagnostics of Molten Salt for its Properties and Structure Online Monitoring

Changzu Zhu[1], Jia Song[1], Xiaorui Xu[1], Chengyu Wang[1], Yang Tong[1], Lve Lin[1], Shaoqiang Guo[2], Wentao Zhou[1], Adrien Couet[3,4], Yafei Wang[1,*]

[1]School of Nuclear Science and Engineering, Shanghai Jiao Tong University, Shanghai 200240, China
[2]Shaanxi Key Laboratory of Advanced Nuclear Energy and Technology, School of Nuclear Science and Technology, Xi'an Jiao Tong University, Xi'an 710049, China
[3]Department of Engineering Physics, University of Wisconsin-Madison, WI 53706, USA
[4]Department of Materials Science and Engineering, University of Wisconsin-Madison, WI 53706, USA

## Abstract

Analogous to the aqueous solution where the pH of the solvent affects its multiple behaviors, the Lewis acidity-basicity of molten salts also greatly influences their thermophysical and thermochemical properties. In the study, we develop ion probes to quantitatively determine the acidity-basicity scale of molten NaCl-$x$AlCl$_3$ ($x$ = 1.5-2.1) salt using in-situ ultra-violet visible (UV-Vis) spectroscopy. With the accumulation of acidity-basicity data of NaCl-AlCl$_3$ molten salt for a variety of compositions, the correlation between the acidity-basicity of salt and its measured fundamental properties are derived. To understand the physical and chemical features controlling the acidity-basicity variations, the structures of NaCl-$x$AlCl$_3$ molten salts with different chemical compositions are investigated in terms of bonded complexes and coordination numbers. The comprehensive understanding of the correlation between composition, acidity-basicity, properties, and structures of molten salt can serve for the full screening and online monitoring of salt melt in extreme environments by simply measuring the salt acidity-basicity as developed in this study.

---

\* Corresponding author.

*E-mail address:* itsme@sjtu.edu.cn

# Introduction

Molten salt plays a crucial role in various advanced energy technologies, for instance, as a thermal energy storage/transfer media for concentrated solar power plants [1, 2], coolant/fuel carrier for Gen-IV molten salt nuclear reactors [3, 4], electrolytes for batteries and pyroprocessing of spent nuclear fuel [5, 6], etc. However, the developments of these technologies are currently limited by a lack of knowledge of the thermophysical and thermochemical properties of molten salts at their operating temperatures. Indeed, properties such as viscosity, diffusivity, redox potential, and corrosivity have a significant impact on thermo-hydraulics, heat transfer, reaction rate, and materials degradation processes. These properties are primarily dictated by the chemistry, structure, and speciation of compounds in molten salt, and the characterization of these properties remains a challenge, especially in complex molten salt systems under extreme environments. Although the techniques for the individual measurement of these properties, e.g., the electrochemical method, have been well developed, seeking an efficient way to synthetically characterize the thermophysical and thermochemical properties of molten salts at different timescales is still necessarily needed.

The pH scale widely applied in aqueous solution to describe the acidity or basicity is closely related to its inherent properties including corrosivity [7], solubility [8], reaction rate [9], etc. It can act as an ideal index for the quantitative evaluation of different properties for aqueous solutions as a whole. The acidity-basicity of molten salt also influences its properties and can be induced to gain knowledge of molten salt thermochemical and thermophysical properties. However, unlike an aqueous solution in which the pH scale is simply defined by the ability of an acid to donate protons, as described by the Bronsted-Lowry theory [10], there exists no quantitative scale of acidity or basicity for the molten salt system. In analogy to the pH scale, oxyanion scales such as $pO^{2-}$, or oxide scales such as $pLi_2O$, have been used to characterize molten salt oxoacidity. Potential-oxoacidity diagrams of Cr, Ni and Fe have been developed using $pLi_2O$ [11]. This approach has had some success, but only in certain restricted conditions because single ion or oxide activities cannot be defined as such, especially in multicomponent mixtures [12]. Thus, a new quantitative acidity-basicity scale, valid across a wide range of molten salt compositions, is still essential for the development of molten salt technologies.

The Lewis theory of acids and bases is one of the most general and is especially appropriate for dealing with non-aqueous and non-protonic systems [13]. In applying the Lewis concept, one envisages the basicity as the ability to share electrons with an acidic solute. This ability can be quantified if a suitable metal ion is selected as an acid probe. The acid probe can be coordinated by a Lewis base and its outer orbitals would be profoundly expanded due to the effects of central-field and symmetry-restricted covalency [14]. This expansion has been measured experimentally in the probe ions of $Tl^+$, $Pb^{2+}$, and $Bi^{3+}$ as the frequency of the 6s → 6p electronic transition [15]. Thus, by measuring the frequency of the 6s → 6p electronic transition of the probe ion in the solvent system using the optical spectrum, such as UV-Vis, it is possible to quantitatively determine the basicity of non-aqueous or non-protonic systems. The measured basicity was called the optical basicity scale and has been successfully demonstrated to extract acidity-basicity properties in molten glass systems [13-17].

The NaCl-$x$AlCl$_3$ molten salts with various Al/Na ratios can melt at low temperatures (about 443 K) and exhibit significant discrepancies on acidity-basicity. It was chosen as the solvent to study the acidity-basicity of the molten salts [18]. Inspired by the establishment of the optical basicity scale in molten glass systems, in this work, we build the acidity-basicity scale for NaCl-$x$AlCl$_3$ molten salt by using the developed ion probes. This is achieved by dissolving the ion probes in molten salts, the probe cations are coordinated by the chlorine atoms of molten NaCl-$x$AlCl$_3$ salts and behave in certain fundamental features which are detected by UV-Vis signatures on the s-p spectra of the ion probe. Based on the frequency obtained on the s-p spectra for different NaCl-$x$AlCl$_3$ salt mixtures, the acidity-basicity scale is determined by a model as already well-developed for molten glass systems [14]. With the accumulation of acidity-basicity data for a variety of NaCl-$x$AlCl$_3$ molten salt mixtures, the correlation between acidity-basicity and salt composition is derived. Considering the great influences of salt acidity-basicity on its properties, the properties including viscosity, diffusivity, and corrosivity for different NaCl-$x$AlCl$_3$ salt mixtures are investigated and their associations with acidity-basicity are explored. To understand the physical and chemical features controlling the acidity-basicity variations with chemical compositions in a molten salt system, the structures of NaCl-$x$AlCl$_3$ molten salts with different compositions are also investigated. The study provides a comprehensive understanding of the correlation between acidity-basicity, properties, and structures of molten salt, which allows the quantification of salt acidity-basicity for being used for the rapid online diagnostics of molten salt on its properties and structures.

**Results**

**Quantitative determination of molten salt acidity-basicity**

The present study focuses on the quantitative determination of acidity-basicity for molten NaCl-$x$AlCl$_3$ salts. In molten chloride salt, one can consider it as an array of its constituent cations and Cl$^-$ ions. The chemical bonding between these species uses much of the negative charge on chlorine and the ability of chlorine to donate negative charge is recognized as the basicity. This ability varies with salt composition and temperature, and the measurement of these changes in electronic charge is the heart of the quantitative determination of molten salt acidity-basicity. For the p-block metal ions in oxidation states two units less than the number of the group to which they belong, for example, Tl$^+$, Pb$^{2+}$, and Bi$^{3+}$, which can receive negative charge from chlorine and experience expansion in the 6s orbital by "nephelauxetic" effect as shown in Fig. 1A. Due to that, the metal ions (Tl$^+$, Pb$^{2+}$, Bi$^{3+}$, etc.,) will undergo enormous changes in the frequency of their absorbing UV band when a small concentration of them is added into molten salt and this has been successfully demonstrated in molten glass system [14]. These metal ions are very sensitive to the changes in the negative charge donation ability of chlorine and thus can be utilized for the acidity-basicity measurement of molten salt by probing the shift on UV frequency.

To achieve the measurement of acidity-basicity for NaCl-$x$AlCl$_3$ molten salts with different compositions, the metal ions including Tl$^+$, Pb$^{2+}$, and Bi$^{3+}$ were dissolved into the salt as ion probes by the dilute addition of TlCl, PbCl$_2$, and BiCl$_3$, respectively. Before the measurement of acidity-basicity for NaCl-$x$AlCl$_3$ molten salt, the selection of ion probe from Tl$^+$, Pb$^{2+}$, and Bi$^{3+}$ was determined by the comparison of UV-Vis spectroscopies of these three metal ions in one same molten salt solute. The metal ion of Tl$^+$ was finally selected for the subsequent measurement because of the peak singularity and less noise in its UV-Vis spectrum (see Fig. S1 in the supplementary document). The in-situ UV-Vis experiments were carried out in a custom spectrum furnace as shown in Fig. 1B. The ultra-violet visible light originating from the light source was guided into the molten salt sample contained in a quartz cuvette through a custom cuvette holder made of 316 L stainless steel. The transport of the light beam from the light source to the spectrometer was achieved by a high-temperature-resistant fiber cable. The UV-Vis spectrum tests were performed for 7 different NaCl-$x$AlCl$_3$ molten salt mixtures at 443 K (all salt mixtures were fully melted). For each salt sample, the UV-Vis spectrum was recorded for 5 times to avoid accidental errors. Fig. 1C shows the representative UV-Vis spectra for NaCl-$x$AlCl$_3$ ($x$ = 1.5-2.1) molten salt mixtures. Other UV-Vis spectra from repeat tests for each molten salt mixture are displayed in Fig. S2 in the supplementary document.

In order to determine the acidity-basicity scale of NaCl-$x$AlCl$_3$ molten salt with different compositions, it is necessary to set up a reference salt whose acidity-basicity scale is defined to be 1. In this study, the NaCl-2.0AlCl$_3$ molten salt mixture is selected as the reference, then the acidity-basicity scale $\Lambda$ can be calculated using equation (1)

$$\Lambda = \frac{v_f - v}{v_f - v_1} \qquad (1)$$

where $v_f$ is the frequency of the free probe ion, which is 55300 cm$^{-1}$ for Tl$^+$ at the free state as reported in the previous literature [14], $v$ is the frequency of the probe ion in the salt medium to be measured, and $v_1$ is the frequency of the system in which the acidity-basicity scale is assumed to be 1.

The frequency of Tl$^+$ was extracted from the UV-Vis absorption spectra. The acidity-basicity scales of each molten salt mixture were calculated using equation (1) and the results of five repeat experiments with the same salt were averaged (see Table S1 in the supplementary document). Fig. 1D shows the variation of the acidity-basicity scale of molten salt with the Al/Na mole ratios in NaCl-AlCl$_3$ salt mixtures. It indicates the acidity-basicity scale decreases with the increasing mole ratio of Al/Na.

With the accumulation of acidity-basicity data for NaCl-AlCl$_3$ molten salt system in a wide variety of compositions, it is significant to consider how the acidity-basicity scale $\Lambda$ is related to the chemical composition for acidity-basicity prediction. In molten salt systems, the neutralization is dependent on the proportion of negative charge each cation neutralizes and the polarizing or electron-attracting property of each cation which is characterized as a moderating parameter, $\gamma$. The relationship between these neutralization-affecting factors and $\Lambda$ for a binary system can be described by a semi-

empirical equation [19] as shown below

$$\Lambda = X_A \times \frac{1}{\gamma_A} + X_B \times \frac{1}{\gamma_B} \tag{2}$$

$$X_A = \frac{ax}{ax + by} \tag{3}$$

$$X_B = \frac{by}{ax + by} \tag{4}$$

where *a* and *b* are charge numbers carried by cations of A and B, respectively. *x:y* is the mole ratio of cation A to cation B. $\gamma_A$ and $\gamma_B$ are the moderating parameters which can be extracted by fitting equation (2) with the acidity-basicity database for different NaCl-AlCl$_3$ molten salt mixtures. Based on the above equation, a multivariate linear fit to the experimental data was performed to obtain the variation of $\Lambda$ as a function of the salt composition. In this study, the $\gamma_{Na}$ and $\gamma_{Al}$ were fitted to be 0.6864 and 1.0808, respectively, with the residual sum of squares value of 8.2997×10$^{-6}$. The parity plot in Fig. 1E shows the predicted $\Lambda$ of different NaCl-AlCl$_3$ molten salt mixtures focused in this study versus the experimental values, from which a very good agreement can be observed. Therefore, based on the fitted moderating parameters $\gamma_{Na}$ and $\gamma_{Al}$, equation (2) can be simplified to equation (5), and be utilized to predict the acidity-basicity scale $\Lambda$ for the NaCl-xAlCl$_3$ molten salt system covering the entire composition range.

$$\Lambda = 1.4569 X_{Na} + 0.9252 X_{Al} \tag{5}$$

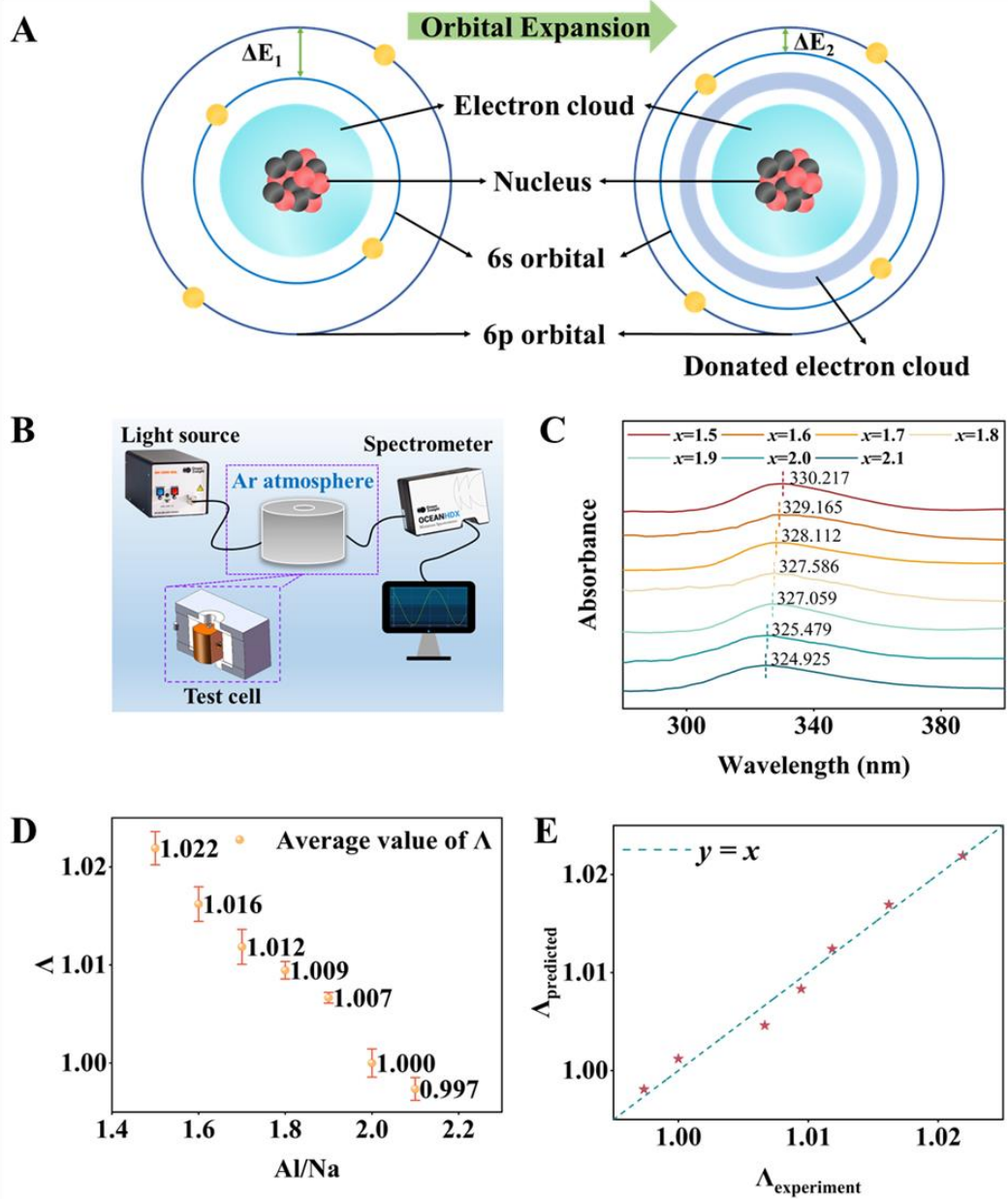

**Fig. 1. The acidity-basicity scale determination of NaCl-$x$AlCl$_3$ molten salt mixture with different Al/Na ratios.**

**(A)** The orbital expansion caused by the nephelauxetic effect of probe ions in molten salts (the donated electron cloud represents the newly generated electron cloud after the ion probe receives the electrons donated by the anions in the melt); **(B)** The schematic of the in-situ UV-Vis measurement for high-temperature molten salt; **(C)** The representative UV-Vis spectra for NaCl-$x$AlCl$_3$ ($x$ = 1.5-2.1) molten salts; **(D)** The calculated acidity-basicity scale of NaCl-$x$AlCl$_3$ ($x$ = 1.5-2.1) molten salt systems based on UV band shift; **(E)** The comparison of predicted and experimental values of acidity-basicity scale.

## Effect of molten salt structure on acidity-basicity scale

The acidity-basicity scale is influenced by the coordination of metal cations and chlorine ions in molten chloride salts. Thus, to develop a robust fundamental understanding of physical and chemical features controlling the acidity-basicity scale variations with chemical compositions, it is necessary to study how the acidity-basicity scale is correlated to the molten salt structure. As an ideal way to investigate the local structure of metal cations in molten salt, deep potential molecular dynamics (DPMD) simulation was employed to provide information on the first shells of metal atom neighbors in terms of coordination number and interatomic distance. The DPMD simulation uses the computational results of the ab initio molecular dynamics (AIMD) simulations based on density functional theory (DFT) as the initial data, machine learning (convolutional neural network) to train a potential function that fits the system of interest, and further applies it to the classical molecular dynamics (MD) simulation. It takes advantage of the accuracy of AIMD calculations and the efficiency of classical MD simulations. The main process of training deep potential (DP) is shown in Fig. 2A, consisting of three parts: training, exploration and labeling. This whole process is repeated until the accuracy of the DP is higher than the set requirement.

The root-mean-square error (RMSE) is commonly used to evaluate the deviation degree of the results predicted by DPMD and DFT simulations, with smaller values indicating a closer computational accuracy of these two methods. Fig. 2B and Fig. 2C show the RMSEs of the energy and force as a function of training steps for NaCl-$x$AlCl$_3$ ($x$ = 1.5-2.1) molten salt system, which are 13.96 meV/atom and 160.85 meV/Å, respectively, when being stable after the training step of around $1\times10^5$. Therefore, the machine learning training step of $1\times10^6$ performed in this study is far sufficient to train the DP model. It is also found the calculated RMSEs of the energy and force are quite small, which is close to or even lower than the values reported previously by Dai et al. [20], implying that the excellent accuracy of the trained DP model in this study. In addition, the trained DP was further examined by comparing the results of energy and force calculated by DPMD and DFT as shown in Fig. 2D and Fig. 2E. In these two figures, the results of energy and force of DFT and DPMD are close to each other, indicating that the trained DP model in this study has achieved a level of accuracy comparable to that of DFT and can be utilized for the subsequent study of molten salt structure.

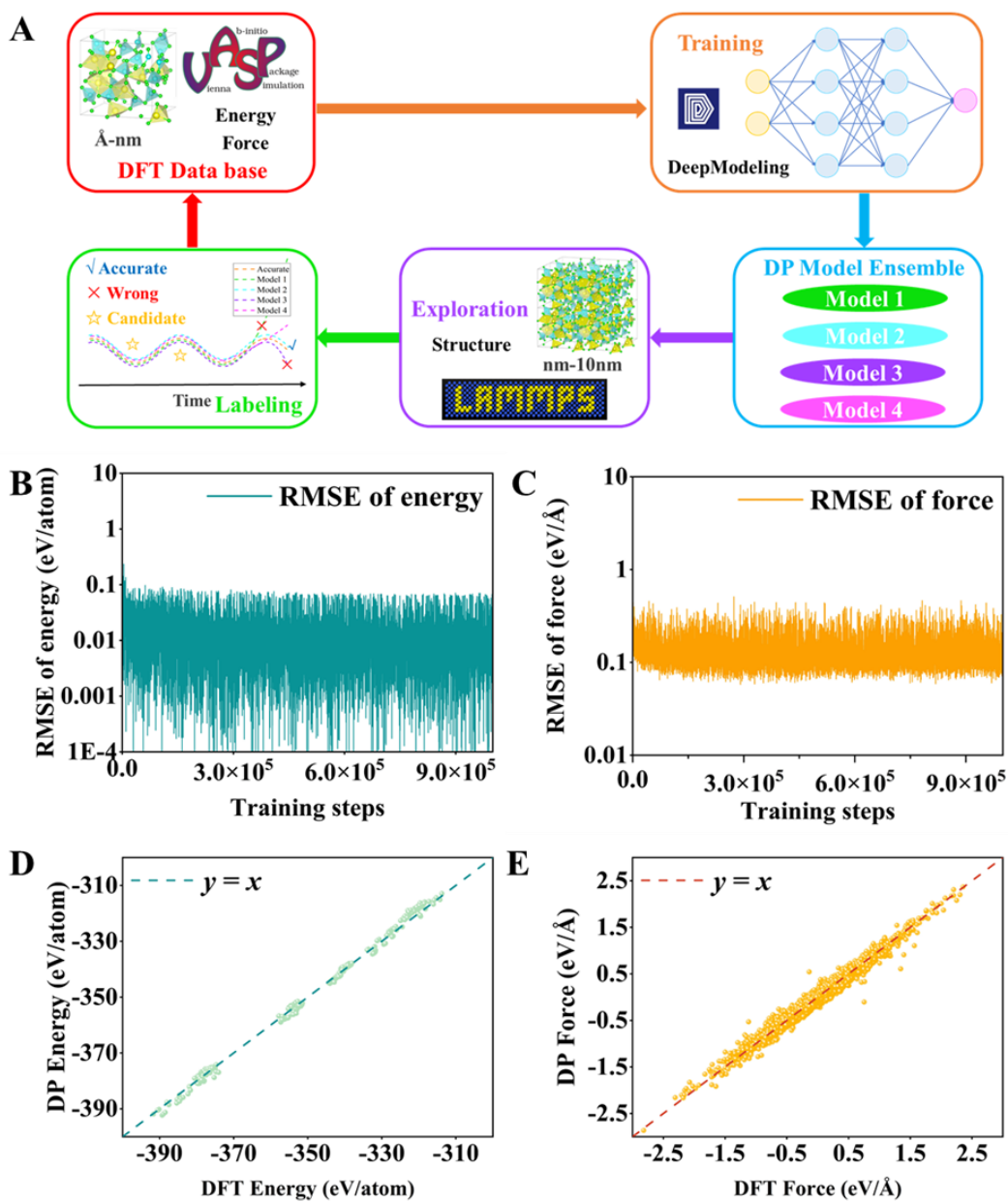

**Fig. 2. Workflow and performance of DP models.**
**(A)** Workflow for constructing the DP model; The RMSEs of energy **(B)** and force **(C)** for the NaCl-$x$AlCl$_3$ ($x$ = 1.5-2.1) molten salt system; The comparison of DFT and DPMD predicted energies **(D)** and forces **(E)** for NaCl-$x$AlCl$_3$ molten salt mixtures.

    The radial distribution function (RDF) has been proven to be an effective way of characterizing the structure of molten salt, with peak positions and integral intensities reflecting structural information such as bond length and coordination number. Fig. 3A and Fig. 3B show the calculated RDFs for NaCl-1.0AlCl$_3$ molten salt mixture at 443 K and each ion pair in the salt mixture by DPMD simulation (RDFs for other NaCl-$x$AlCl$_3$ molten salt mixtures are displayed in Fig. S3 in the supplementary document). The RDFs calculated by AIMD are also given in Fig. 3A and Fig. 3B as a comparison from which a good agreement can be observed. Additionally, the experimental RDF [21] of

the Al-Cl ion pair in the NaCl-1.0AlCl$_3$ molten salt mixture shown in Fig. 3C exhibits a good consistency on the peak position with the RDF calculated by DPMD as well. This allows more confidence in studying the local structures of NaCl-$x$AlCl$_3$ molten salt mixtures by DPMD simulation.

Based on the calculated RDFs, the detailed bond length and coordination number were extracted and summarized in

Table **1**. It can be concluded that the changes in the bond length and coordination number (CN) are very limited in the concentration range focused in this study. In NaCl-AlCl$_3$ molten salt, Al atoms exist in the forms of either $AlCl_4^-$ or $Al_2Cl_7^-$ ion clusters [22]. The CN number of around 4 between Al and Cl as shown in

Table **1** indicates that the $AlCl_4^-$ ion cluster dominates in all the NaCl-AlCl$_3$ molten salt mixtures in this study.

Table **1** also shows the CN of Al-Cl decreases slightly with the rise of Al/Na ratio. Considering that the calculated CN between Al and Cl from RDFs is an average of the diverse coordination environment of an Al atom with the Cl atoms nearby in the melt, the variation of the CN of Al-Cl should be due to the proportion change of $AlCl_4^-$ and $Al_2Cl_7^-$ in the salt melt. Since the CN of Al-Cl in $Al_2Cl_7^-$ cluster is 3.5, the decrease of CN with the increase of Al-Na ratio is a result of the increasing proportion of $Al_2Cl_7^-$ cluster through the equilibrium reaction in equation (6)

$$AlCl_4^- + AlCl_3 = Al_2Cl_7^- \qquad (6)$$

**Table 1. The bond length and coordination number of NaCl-$x$AlCl$_3$ ($x$ = 1.5-2.1) melts.**
The $r_{\text{A-B}}$ and CN (A-B) represent the bond length and the coordination number between A atom and B atom, respectively.

| Al/Na | 1 | 1.5 | 1.6 | 1.7 | 1.8 | 1.9 | 2 | 2.1 |
|---|---|---|---|---|---|---|---|---|
| $r_{\text{Al-Cl}}$ | 2.15 | 2.15 | 2.13 | 2.13 | 2.13 | 2.13 | 2.13 | 2.13 |
| $r_{\text{Na-Cl}}$ | 2.87 | 2.89 | 2.89 | 2.85 | 2.85 | 2.89 | 2.87 | 2.83 |
| $r_{\text{Cl-Cl}}$ | 3.55 | 3.53 | 3.53 | 3.55 | 3.55 | 3.55 | 3.55 | 3.55 |
| CN(Al-Cl) | 4.0012 | 4.0011 | 4.0007 | 4.0004 | 3.9993 | 3.9995 | 3.9980 | 3.9974 |

The above study shows the proportion of $Al_2Cl_7^-$ cluster in NaCl-AlCl$_3$ molten salt mixture changes with the salt composition, which might be the main factor leading to the variation of the salt acidity-basicity scale. The structure of $Al_2Cl_7^-$ displayed in the inset of Fig. 3D shows that the two Al atoms in an $Al_2Cl_7^-$ cluster are connected by a bridging Cl atom called corner-sharing Cl. By quantifying the corner-sharing Cl in the salt melt using a self-written Javascript based on Sun's report [23], the proportion of $Al_2Cl_7^-$ cluster can be assessed and the influence of molten salt structure on its acidity-basicity scale can be further explored. As shown in Fig. 3D, the acidity-basicity scale exhibits an obvious decreasing trend as the percentage of corner-sharing Cl increases. It indicates that the NaCl-AlCl$_3$ molten salt mixture shows a stronger Lewis acidity with the increasing proportion of Al$_2$Cl$_7^-$ (or the corner-sharing Cl) in the melt. This finding is also consistent with the previous report [22] on the NaCl-*x*AlCl$_3$ molten salt system. In Fig. 3D, it is noted that the structure of NaCl-1.8AlCl$_3$ (Λ=1.009) melt slightly deviates from the above rule, which should be owing to the numerical uncertainty from the DPMD simulation, but it does not affect the validity of the above conclusion.

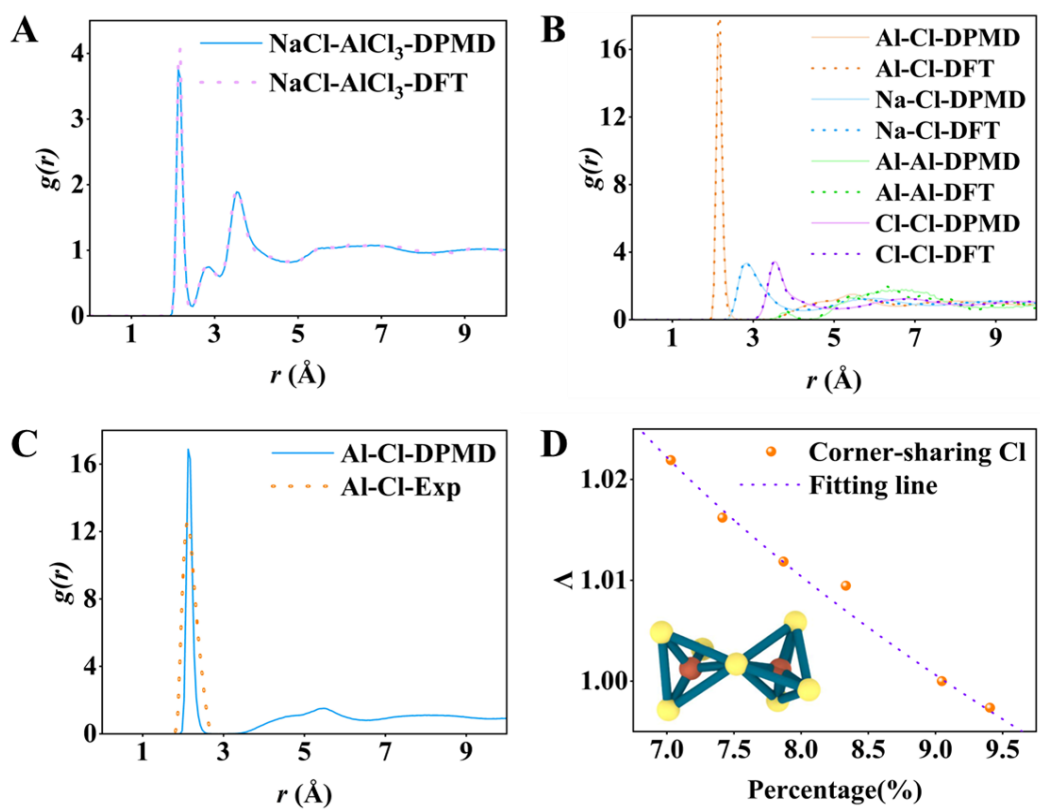

**Fig. 3. The structure analysis of NaCl-*x*AlCl$_3$ molten salt systems.**
**(A)** A comparison of total RDFs of NaCl-1.0AlCl$_3$ melts at 443 K calculated by AIMD and DPMD simulations; **(B)** RDFs of Al-Cl, Na-Cl, Al-Al and Cl-Cl ion pair for NaCl-1.0AlCl$_3$ molten salt mixture at 443 K; **(C)** A comparison between RDF of Al-Cl ion pair obtained by DPMD and experiment in NaCl-1.0AlCl$_3$ molten salt mixture at 443 K; **(D)** The relationship between the salt acidity-basicity scale and the percentage of corner-sharing Cl in NaCl-AlCl$_3$ molten salt mixture. The inset shows the structure of Al$_2$Cl$_7^-$ where the yellow and orange balls represent Cl and Al atoms, respectively.

## Correlation of acidity-basicity scale with molten salt properties

The variation of thermal transport properties of NaCl-*x*AlCl$_3$ molten salt mixtures with different acidity-basicity scales was analyzed using DPMD simulations. The viscosity of molten salts is a significant parameter for evaluating the internal flow rate of molten salts, which is related to the diffusion behavior of solute ions and mass transfer processes in solution, and it can be calculated via the Green−Kubo (GK) approach [24]:

$$\eta = \frac{V}{k_B T} \int_0^\infty \langle P_\alpha(t) \cdot P_\alpha(0) \rangle \mathrm{d}t \tag{7}$$

where $V$ is the volume, $K_B$ is the Boltzmann constant, $T$ is the temperature, the angle bracket denotes the average overall time, and $P_\alpha$ represents the element $\alpha$ of the pressure tensor. In this part, the average running integrals of multiple independent trajectories are utilized instead of the simulation of long-time trajectories to minimize the effect of statistical noise [25]. Viscosity calculations were carried out in the NVT ensemble at 443 K for 4 ns using five different trajectories (each trajectory is 0.8 ns long). On the basis of each trajectory, the viscosity was calculated via equation (7) and the calculated results for NaCl-1.5AlCl$_3$ molten salt mixture are shown in Fig. 4A (the results for other salt mixtures are shown in Fig. S4). The viscosities obtained from these five trajectories were averaged as the final value for the viscosity to be studied. Fig. 4B shows the averaged viscosities for different NaCl-AlCl$_3$ molten salt mixtures in which the values are found to be in the range from 1.636 to 2.471 mPa·s. Comparing the calculated viscosities of NaCl-1.5AlCl$_3$ and NaCl-1.8AlCl$_3$ molten salt mixtures with the reported experiment values [26], a good agreement can be observed. However, the change of viscosity versus acidity-basicity scale shown in Fig. 4B does not exhibit an obvious correlation between these two parameters.

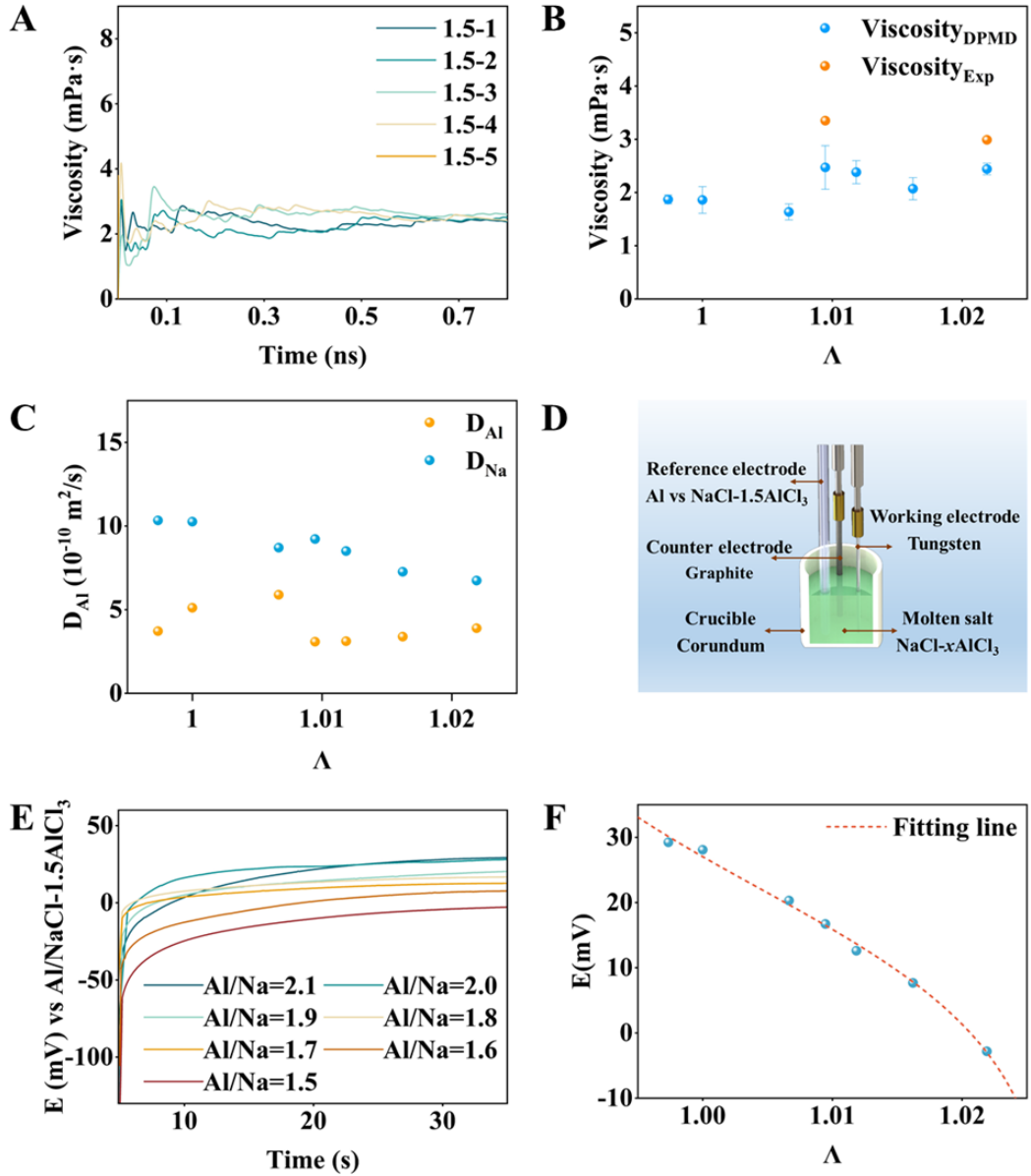

**Fig. 4. The correlation of acidity-basicity scale of NaCl-$x$AlCl$_3$ molten salt mixtures with its fundamental properties.**
**(A)** Viscosities of NaCl-1.5AlCl$_3$ melts calculated by different trajectories, 1.5-1 to 1.5-5 represent the five trajectories used to calculate viscosities; **(B)** The relationship between the viscosity of molten salt and its acidity-basicity scale; **(C)** The variations of self-diffusion coefficients of Al and Na atoms in NaCl-$x$AlCl$_3$ ($x$ = 1.5-2.1) melts with the salt acidity-basicity scale; **(D)** Schematic of the high temperature molten salt electrochemical test cell; **(E)** Open circuit potential evolution of NaCl-$x$AlCl$_3$ molten salt systems with different Al/Na ratios at 443 K after electrodepositing Al on working electrode; **(F)** Variation of the reduction potential of Al with the acidity-basicity scale of NaCl-AlCl$_3$ molten salt mixtures.

In addition, the self-diffusion coefficient which reflects the ability of atoms to diffuse in molten salt, is also analyzed according to Einstein's expression (equation (8)) [27] in this work.

$$D = \lim_{t \to \infty} \frac{1}{6} \frac{d\langle MSD \rangle}{dt} \quad (8)$$

where $t$ is the time elapsed and the MSD denotes the mean square displacement, which is used to measure the deviation of the position of the particle after moving over time with respect to the reference position, and it can be calculated as follows:

$$\text{MSD} = \langle |\delta r_i(t)|^2 \rangle = \frac{1}{N} \langle \sum_i |r_i(t) - r_i(0)|^2 \rangle \quad (9)$$

where $\delta r_i(t)$ is the displacement of the atoms in time $t$, $r_i(t)$ is the position at the time of $t$ while $r_0(t)$ is the position at the time of 0. The calculated MSDs of Al and Na ions as a function of elapsed time (up to 1 ns) in NaCl-$x$AlCl$_3$ melts with different Al/Na ratios are displayed in Fig. S5 in the supplementary document. Based on equation (8), the self-diffusion coefficient can be deduced from the slope of the MSD versus time. Fig. 4C shows the calculated self-diffusion coefficients of Al and Na ions in different NaCl-AlCl$_3$ molten salt mixtures. It is obvious from Fig. 4C that the self-diffusion coefficients of Na ions are greater than that of Al ions. In NaCl-$x$AlCl$_3$ molten salt mixtures, Al exists in the form of either $AlCl_4^-$ or $Al_2Cl_7^-$ ion cluster. The cluster hinders the mass transfer of Al ions in the salt and which might be the main reason for the lower diffusivity of Al. From Fig. 4C, no obvious variation trend between the salt acidity-basicity scale and diffusion coefficient was observed, indicating that the diffusion coefficient has no direct correlation with the salt acidity-basicity scale.

Reduction potential is a key electrochemical parameter that determines the type of chemical reaction in molten salt. It can be used to guide electrolytic extraction and learn salt's corrosivity. As the ion with the strongest oxidation ability in NaCl-$x$AlCl$_3$ molten salt mixtures, the reduction potential of Al ion determines the salt's corrosivity. To study the correlation of salt acidity-basicity with its corrosivity, the reduction potentials of Al ions in NaCl-$x$AlCl$_3$ melts were measured by a customized high-temperature molten salt electrochemical test cell as shown in Fig. 4D. The measurement of the reduction potential was achieved by electrodepositing Al metal on tungsten electrode from molten salt and measuring the open circuit potential (OCP) evolution between the working electrode and reference electrode. The stabilized OCP value corresponds to the reduction potential of Al ions in NaCl-$x$AlCl$_3$ molten salt mixtures. The OCP evolutions with time for different NaCl-$x$AlCl$_3$ mixtures are shown in Fig. 4E from which it can be seen the stabilized OCP becomes more positive with the increase of the Al/Na ratio. The stabilized OCP values for different NaCl-$x$AlCl$_3$ molten salt mixtures were collected as the reduction potentials. Based on the above measured acidity-basicity scale for each corresponding NaCl-$x$AlCl$_3$ molten salt mixture, the relationship between the reduction potential and acidity-basicity is plotted in Fig. 4F.

Combining the Nernst equation[28] and the fitted correlation equation between the acidity-basicity scale and the composition of molten salt, the relationship between reduction potential $E$ and acidity-basicity scale can be built by

$$E = E_0 - \frac{RT}{nF} \times \ln\left(\frac{p\Lambda + a}{q\Lambda + b}\right) \qquad (10)$$

where $E_0$ is the standard potential, $R$ is the gas constant, $T$ is the temperature, $n$ is the number of transferred electrons, $F$ is the Faraday constant, and $p$, $q$, $a$, and $b$ are the fitting parameters. Based on the measured acidity-basicity scale and reduction potential for different NaCl-$x$AlCl$_3$ molten mixtures, the equation (10) can be utilized to fit the experimental data as displayed in Fig. 4F. The obtained $p$, $q$, $a$, and $b$ are -6.32465, 6.3335, 6.49875 and -6.21421, respectively. As shown in Fig. 4F, the measured plots are well distributed around the fitting line, indicating the equation (10) can well describe the variation of reduction potential with the acidity-basicity scale of molten salt. The reduction potential of Al gradually increases with the decrease of the acidity-basicity scale and it should be attributed to the growth in the concentration of Al$_2$Cl$_7^-$, which is the main carrier of electrochemical reactions in Lewis acidic NaCl-$x$AlCl$_3$ melts [29]. According to the above relationship of the acidity-basicity scale with molten salt composition and the correlation of reduction potential with acidity-basicity scale, the reduction potential of Al in the NaCl-$x$AlCl$_3$ system with a given Al/Na ratio can be predicted directly and the salt corrosivity can be learned accordingly as well.

## Discussion

Analogous to the aqueous solution where the pH of solvent is closely related to its multiple inherent properties, there should also be a tight correlation between the acidity-basicity scale of molten salt and its properties. The measurement of salt acidity-basicity can be one of the most simple and efficient ways to gain knowledge of molten salt thermochemical and thermophysical properties if the correlations are unveiled. This is crucial for the applications of molten salt technologies since it can help learn how the salt changes as a whole, serving for the online monitoring on the variations of molten salt contents and properties. For instance, if the correlation between the salt acidity-basicity scale and composition is known, the concentrations of corrosion products in molten salt can be extracted through the measurement of acidity-basicity and the in-situ corrosion monitoring in the molten salt environment can be further achieved. Similarly, if the correlation between acidity-basicity and salt corrosivity is revealed, the subsequent corrosion attack of material in molten salt can be priorly estimated. Therefore, the comprehensive study of the correlation between acidity-basicity, properties, and structures of molten salt allows the quantification of salt acidity-basicity for being used for the rapid online diagnostics of molten salt on its properties and structures.

Unlike an aqueous solution in which the pH scale is simply defined by the ability of an acid to donate protons, the quantification for the acidity-basicity scale of molten salt system is complex and the relevant studies are rare [30]. Inspired by the establishment and application of the optical basicity in molten glass, this study focused on NaCl-$x$AlCl$_3$ molten salt system and quantified the acidity-basicity scale. Different from the optical basicity of molten glass, the acidity-basicity scale determined for molten salt in this study chose the reference salt whose acidity-basicity scale is defined to be 1. The selection of reference salt for the determination of the acidity-basicity scale can be arbitrary. It means the acidity-basicity scale for the same salt system could change if a different reference salt is chosen. By studying the acidity-basicity scales of different NaCl-AlCl$_3$ molten salt mixtures, the relationship between the acidity-basicity scale and the composition of NaCl-$x$AlCl$_3$ molten salts has been built through a moderating parameter, γ. This built relationship can be used to learn the composition variations of molten salt if the acidity-basicity scale is known. Therefore, the primary task for developing and applying the acidity-basicity scale of molten salt is to accumulate data in various systems to establish the database and derive γ-value for each element. The acidity-basicity scale measured by ion probes in this study is based on an assessment of the charge density carried by anions [13]. Thus the correlation between the value of γ of each element in the molten salt system and its electronegativity deserves further investigation [16]. In addition, analyzing the electronic structure of molten salts using Bader charge analysis [31] and electron localization function [32] in DFT simulations may help to perceive the nature of the acidity-basicity scale more clearly.

To develop a robust fundamental understanding of physical and chemical features controlling the acidity-basicity scale variations with chemical compositions, the influence of molten salt structure on its acidity-basicity scale is studied. It is found the concentration of $Al_2Cl_7^-$ linked by corner-sharing Cl atoms increases with the reduction of the acidity-basicity scale, Λ. It has been widely reported that the cations usually form networks with Cl or F atoms and change the structure and properties of the melt in the molten salt environment [33, 34]. The study on the correlations of molten salt structure and acidity-basicity scale can be utilized for the online monitoring of changes in melt structure through the simple measurement of the acidity-basicity scale.

The thermal transport properties of molten salts appear to be insensitive to changes in the acidity-basicity scale and no uniform law is discovered from the data in the present work. It may be attributed that the viscosity and self-diffusion coefficient change only slightly in the range of study, but the DPMD simulation cannot reflect this small change accurately as a result of the statistical error. In terms of the electrochemical properties, the relationship between the acidity-basicity scale and reduction potential is well fitted. The reduction potential of the least thermodynamically stable cation in molten salt can reflect its corrosivity. Based on the fitted relationship between the salt acidity-basicity scale and reduction potential, the corrosivity of molten salt can be derived. Grasping the change of reduction potential with the acidity-basicity scale is beneficial to understanding and mitigating the corrosion behavior of materials more profoundly. The correlation study between acidity-basicity and properties allows the quantification of salt acidity-basicity to be used for the rapid online diagnostics of molten salt properties if these correlations are revealed for more kinds of properties. However, an interesting question is whether the properties of molten salts with different components will be similar when they have equal Λ. If the above assumptions hold, the acidity-basicity scale will be as universal as pH. It can directly compare the overall properties of molten salts with different components and temperatures by comparing their acidity-basicity, and even establish a brand-new evaluation system of molten salt based on it. However, it should be noted that this brand-new evaluation system should have a universal reference salt selected and accepted by the whole community.

## Materials and Methods

### Experimental materials and salt melt preparation

Before in situ UV-Vis spectroscopy and electrochemical experiments, NaCl-$x$AlCl$_3$ ($x$ = 1.5-2.1) salts were prepared by anhydrous aluminum chloride (98.0%) and sodium chloride (99.9%) purchased from Shanghai Aladdin Biochemical Technology Co., Ltd. The salts were weighted by a precision balance with an accuracy of $10^{-5}$ g (Mettler Toledo, MS 205DU/A). 80 grams of each proportion of NaCl-$x$AlCl$_3$ salt was prepared and mixed fully to ensure that the mixture was uniform. About 5 g and 50 g samples were used for each in situ UV-Vis spectroscopy and electrochemical experiment, respectively. The temperature of the furnace was calibrated by thermocouple with a temperature accuracy of ±1.5 °C, and the mixed powder was heated and melted at 443 K.

### In-situ ultraviolet-visible absorption spectroscopy

The in-situ ultraviolet-visible absorption spectrometer mainly consists of a light source, a test cell, a spectrometer, a connecting optical path (optical fiber), etc. The in-situ ultraviolet-visible absorption spectrometer used in this work is shown in Fig. 1B, and the sample cuvette is located in the furnace in a glove box with an Argon atmosphere. The furnace was used to provide the high temperature required for melting the salt while the inert glovebox was utilized to protect the salt sample from water and oxygen interference. The in situ ultraviolet-visible light generated by the deuterium tungsten lamp (Ocean Optics (SHANGHAI) Co., Ltd., DH-2000-BAL) was led out by a high temperature resistant optical fiber (Ocean Optics (SHANGHAI) Co., Ltd.). The high temperature resistance of the optical fiber was achieved by the gold coating on the fiber. The light passes the quartz cuvette carrying molten salt through the customized collimator and the cuvette holder made of 316 L stainless steel. The sample can absorb specific wavelengths of light because of the transition of the outer electrons, and the relevant information can be transmitted from the optical fiber into the spectrometer (Ocean Optics (SHANGHAI) Co., Ltd., OCEAN-HDX-XR) and analyzed.

### Electrochemical apparatus and electrodes

All the electrochemical experiments were performed in a glove box filled with argon atmosphere with water and oxygen concentrations less than 0.1 ppm. Open circuit potential (OCP) was performed to explore the reduction potentials of Al ions and Al clusters in different NaCl-AlCl$_3$ molten salt mixtures at 443K by Gamry Instruments and Gamry Framework software. Once the molten salt systems were stable, a current of 25-50 mA was applied to the working electrode for 5 seconds and the OCP between the working electrode and the reference electrode was measured and recorded. The stabilized OCP value corresponds to the reduction potential of Al ions in NaCl-$x$AlCl$_3$ molten salt mixtures. As shown in Fig. 4D, a corundum crucible was selected as the sample container, and inert tungsten wire and graphite rod ($\Phi$ = 1.0 mm, 99.98%) were used as the working electrode and counter electrode, respectively. An aluminum wire ($\Phi$ = 0.9 mm, 99.99%) was inserted into the NaCl-1.5AlCl$_3$ melts in the NMR tube and used as a reference electrode. All the electrodes were ultrasonically cleaned in deionized water for 20 min and dried in an oven before any test.

**Computational details**

In this study, deep potential molecular dynamics (DPMD) simulation was adopted to investigate the structure and properties of NaCl-$x$AlCl$_3$ melts. The main process for DPMD can be divided into three parts. To begin with, ab initio molecular dynamics (AIMD) simulations were performed to collect information needed for DPMD such as structures, force and energy. Then, a deep potential model is trained by using the above dataset from AIMD simulations. Finally, MD simulations were carried out using the trained DP model to predict the structure and properties of the NaCl-$x$AlCl$_3$ melts.

All the AIMD simulations were performed within the Vienna Ab initio Simulation Package (VASP) [35-37]. The projector augmented-wave (PAW) method [38, 39] was employed to characterize the interactions between the nucleus and electrons. The Perdew-Burke-Enrzerhof (PBE) [40] was used to describe the exchange-correlation effect. The initial configurations were obtained by Packmol [41] code according to experimental densities, and the numbers of atoms of each configuration are listed in Table S2. Moreover, the DFT-D3 method [42] was used to deal with the dispersion interaction. All AIMD simulations were performed using a 1 × 1 × 1 k-point mesh, and the cut-off energy was set to 420 eV. The steps of AIMD simulations were as follows. Firstly, the initial configuration was equilibrated for 4 ps in an NVT ensemble at 943K. Secondly, the obtained structure was quenched to 443 K. Finally, the system employed 20 ps NVT simulations at 443 K after volume optimization.

The deep potential generator (DP-GEN) package [43] was adopted in the active learning procedure. As shown in Fig. 2A, the concurrent learning process of the DP-GEN consists of three main parts: training, exploration, and labeling [44]. In the first iteration, four deep potential (DP) models were trained with the AIMD results as the initial training data set. The maximum standard deviation of the predicted atomic forces for these four DP models was calculated, and the configurations explored by MD simulations are classified as accurate sets, candidate sets and wrong sets according to its value. Then, the accurate candidate structures for DFT simulations were added to the training dataset for the next iteration. The DP-GEN will keep looping the above three steps until the accuracy of the obtained DP is greater than 99.5% [45]. The DeePMD-kit package [46] was applied to train the DPs of NaCl-$x$AlCl$_3$ melts. During the training of DPs, the sizes of the embedding network and fitting network were set to {25, 50, 100} and {240, 240, 240}, respectively. Considering the balance between computational accuracy and computational resource consumption, the radial cutoff parameter and the smooth cutoff parameter were set to 6.0 Å and 0.5 Å. As regards the prefactors of the energy and force terms in the loss function, which were changed from 0.002 to 1 and 1000 to 1, respectively.

The trained DPs were used to carry out molecular dynamics (MD) simulations with the LAMMPs [47] software via the interface provided by the DeePMD-kit package. The initial configurations were generated by Packmol code according to experimental densities, and the numbers of atoms of each configuration are listed in Table S2. Firstly, the initial structures were optimized in the NPT ensemble for 0.5 ns to equilibrium volume. Subsequently, the optimized supercells were optimized in the NVT ensemble for 1.5 ns, while the radial distribution function (RDF) and mean square displacement (MSD) of the melts were analyzed using the last 1 nm of these trajectories. Furthermore, 4-ns DPMD simulations were performed in the NVT ensemble to calculate the viscosities.

## Acknowledgments

The authors gratefully acknowledge the financial support provided for project number 12305395 by the National Natural Science Foundation of China and project number 21ZR1435400 by the Science and Technology Commission of Shanghai Municipality. The computations in this paper were run on the Siyuan-1 cluster supported by the Center for High Performance Computing at Shanghai Jiao Tong University.

## Author contributions:

**Conceptualization:** Changzu Zhu, Yafei Wang;
**Methodology:** Changzu Zhu, Jia Song;
**Investigation:** Changzu Zhu, Chengyu Wang, Yafei Wang;
**Visualization:** Changzu Zhu, Yang Tong, Lve Lin;
**Supervision:** Yafei Wang, Wentao Zhou;
**Writing—original draft:** Changzu Zhu
**Writing—review & editing:** Changzu Zhu, Xiaorui Xu, Yafei Wang

## Reference

[1] G. Alva, Y. Lin, G. Fang, An overview of thermal energy storage systems, Energy 144 (2018) 341-378.
[2] E. González-Roubaud, D. Pérez-Osorio, C. Prieto, Review of commercial thermal energy storage in concentrated solar power plants: Steam vs. molten salts, Renewable and sustainable energy reviews 80 (2017) 133-148.
[3] J. Serp, M. Allibert, O. Beneš, S. Delpech, O. Feynberg, V. Ghetta, D. Heuer, D. Holcomb, V. Ignatiev, J.L. Kloosterman, The molten salt reactor (MSR) in generation IV: overview and perspectives, Progress in Nuclear Energy 77 (2014) 308-319.
[4] Y. Wang, C. Zhu, M. Zhang, W. Zhou, Molten salt reactors, Nuclear Power Reactor Designs,

Elsevier2024, pp. 163-183.

[5] V. Giordani, D. Tozier, H. Tan, C.M. Burke, B.M. Gallant, J. Uddin, J.R. Greer, B.D. McCloskey, G.V. Chase, D. Addison, A Molten Salt Lithium–Oxygen Battery, Journal of the American Chemical Society 138(8) (2016) 2656-2663.

[6] H. Liu, X. Zhang, S. He, D. He, Y. Shang, H. Yu, Molten salts for rechargeable batteries, Materials Today 60 (2022) 128-157.

[7] S. Thomas, N. Birbilis, M. Venkatraman, I. Cole, Corrosion of zinc as a function of pH, Corrosion, The Journal of Science and Engineering 68(1) (2012) 015009-1-015009-9.

[8] S. Li, S. Wong, S. Sethia, H. Almoazen, Y.M. Joshi, A.T. Serajuddin, Investigation of solubility and dissolution of a free base and two different salt forms as a function of pH, Pharmaceutical research 22 (2005) 628-635.

[9] S. Nangia, B.J. Garrison, Reaction rates and dissolution mechanisms of quartz as a function of pH, The Journal of Physical Chemistry A 112(10) (2008) 2027-2033.

[10] G.B. Kauffman, The Bronsted-Lowry acid base concept, Journal of Chemical Education 65(1) (1988) 28.

[11] S. Delpech, C. Cabet, C. Slim, G.S. Picard, Molten fluorides for nuclear applications, Materials Today 13(12) (2010) 34-41.

[12] E.A. Guggenheim, The conceptions of electrical potential difference between two phases and the individual activities of ions, The Journal of Physical Chemistry 33(6) (2002) 842-849.

[13] J.A. Duffy, A review of optical basicity and its applications to oxidic systems, Geochimica et Cosmochimica Acta 57(16) (1993) 3961-3970.

[14] J. Duffy, M. Ingram, Establishment of an optical scale for Lewis basicity in inorganic oxyacids, molten salts, and glasses, Journal of the American Chemical Society 93(24) (1971) 6448-6454.

[15] J. Duffy, M. Ingram, Use of thallium (I), lead (II), and bismuth (III) as spectroscopic probes for ionic–covalent interaction in glasses, The Journal of chemical physics 52(7) (1970) 3752-3754.

[16] J. Duffy, M.D. Ingram, An interpretation of glass chemistry in terms of the optical basicity concept, Journal of Non-Crystalline Solids 21(3) (1976) 373-410.

[17] J.A. Duffy, A common optical basicity scale for oxide and fluoride glasses, Journal of non-crystalline solids 109(1) (1989) 35-39.

[18] J. Chen, Y. Zhong, Y. Liu, L. Zhang, M. Li, W. Han, Z. Chai, W. Shi, Electrochemical Behaviour and Chemical Species of Sm(II) in $AlCl_3$-NaCl with Different Lewis Acidity, Chemistry – A European Journal 28(42) (2022) e202200443.

[19] J. Duffy, M. Ingram, Optical basicity—IV: Influence of electronegativity on the Lewis basicity and solvent properties of molten oxyanion salts and glasses, Journal of Inorganic and Nuclear Chemistry 37(5) (1975) 1203-1206.

[20] F.-Z. Dai, B. Wen, H. Xiang, Y. Zhou, Grain boundary strengthening in $ZrB_2$ by segregation of W: Atomistic simulations with deep learning potential, Journal of the European Ceramic Society 40(15) (2020) 5029-5036.

[21] Y.S. Badyal, D.A. Allen, R.A. Howe, The structure of liquid $AlCl_3$ and structural modification in $AlCl_3$-MCl (M=Li, Na) molten salt mixtures, Journal of Physics: Condensed Matter 6(47) (1994) 10193.

[22] G. Mamantov, G. Torsi, Potentiometric study of the dissociation of the tetrachloroaluminate ion in molten sodium chloroaluminates at 175-400. deg, Inorganic Chemistry 10(9) (1971) 1900-1902.

[23] M. Sun, C. Han, Y. Lan, Hierarchical fivefold symmetry in CuZr metallic glasses, Journal of Non-Crystalline Solids 555 (2021) 120548.

[24] Y. Zhang, A. Otani, E.J. Maginn, Reliable Viscosity Calculation from Equilibrium Molecular Dynamics Simulations: A Time Decomposition Method, Journal of Chemical Theory and Computation 11(8) (2015) 3537-3546.

[25] M. Mouas, J.-G. Gasser, S. Hellal, B. Grosdidier, A. Makradi, S. Belouettar, Diffusion and viscosity of liquid tin: Green-Kubo relationship-based calculations from molecular dynamics simulations, The Journal of Chemical Physics 136(9) (2012).

[26] G.J. Janz, R. Tomkins, C. Allen, J. Downey Jr, G. Garner, U. Krebs, S.K. Singer, Molten salts: volume 4, part 2, chlorides and mixtures—electrical conductance, density, viscosity, and surface tension data, Journal of Physical and Chemical Reference Data 4(4) (1975) 871-1178.

[27] D. Frenkel, B. Smit, M.A. Ratner, Understanding Molecular Simulation: From Algorithms to Applications, Physics Today 50(7) (1997) 66-66.

[28] A.J. Bard, L.R. Faulkner, H.S. White, Electrochemical methods: fundamentals and applications, John Wiley & Sons2022.

[29] J. Xu, J. Zhang, Z. Shi, Extracting aluminum from aluminum alloys in AlCl3-NaCl molten salts, High Temperature Materials and Processes 32(4) (2013) 367-373.

[30] L. Sharpless, J. Moon, D. Chidambaram, Optical Basicity for Understanding the Lewis Basicity of Chloride Molten Salt Mixtures, The Journal of Physical Chemistry Letters 15(20) (2024) 5529-5534.

[31] G. Henkelman, A. Arnaldsson, H. Jónsson, A fast and robust algorithm for Bader decomposition of charge density, Computational Materials Science 36(3) (2006) 354-360.

[32] A.D. Becke, K.E. Edgecombe, A simple measure of electron localization in atomic and molecular systems, The Journal of chemical physics 92(9) (1990) 5397-5403.

[33] D. Corradini, P.A. Madden, M. Salanne, Coordination numbers and physical properties in molten salts and their mixtures, Faraday discussions 190 (2016) 471-486.

[34] V. Dracopoulos, B. Gilbert, G. Papatheodorou, Vibrational modes and structure of lanthanide fluoride–potassium fluoride binary melts LnF 3–KF (Ln= La, Ce, Nd, Sm, Dy, Yb), Journal of the Chemical Society, Faraday Transactions 94(17) (1998) 2601-2604.

[35] G. Kresse, J. Hafner, Ab initio molecular-dynamics simulation of the liquid-metal–amorphous-semiconductor transition in germanium, Physical Review B 49(20) (1994) 14251.

[36] G. Kresse, J. Furthmüller, Efficiency of ab-initio total energy calculations for metals and semiconductors using a plane-wave basis set, Computational materials science 6(1) (1996) 15-50.

[37] G. Kresse, J. Furthmüller, Efficient iterative schemes for ab initio total-energy calculations using a plane-wave basis set, Physical review B 54(16) (1996) 11169.

[38] P.E. Blöchl, Projector augmented-wave method, Physical Review B 50(24) (1994) 17953-17979.

[39] G. Kresse, D. Joubert, From ultrasoft pseudopotentials to the projector augmented-wave method, Physical Review B 59(3) (1999) 1758-1775.

[40] J.P. Perdew, K. Burke, M. Ernzerhof, Generalized gradient approximation made simple, Physical review letters 77(18) (1996) 3865.

[41] L. Martínez, R. Andrade, E.G. Birgin, J.M. Martínez, PACKMOL: A package for building initial configurations for molecular dynamics simulations, Journal of Computational Chemistry 30(13) (2009) 2157-2164.

[42] S. Grimme, J. Antony, S. Ehrlich, H. Krieg, A consistent and accurate ab initio parametrization of density functional dispersion correction (DFT-D) for the 94 elements H-Pu, The Journal of Chemical Physics 132(15) (2010).

[43] Y. Zhang, H. Wang, W. Chen, J. Zeng, L. Zhang, H. Wang, E. Weinan, DP-GEN: A concurrent

learning platform for the generation of reliable deep learning based potential energy models, Computer Physics Communications 253 (2020) 107206.

[44] R. He, H. Wu, L. Zhang, X. Wang, F. Fu, S. Liu, Z. Zhong, Structural phase transitions in SrTi O 3 from deep potential molecular dynamics, Physical Review B 105(6) (2022) 064104.

[45] J. Huang, L. Zhang, H. Wang, J. Zhao, J. Cheng, Deep potential generation scheme and simulation protocol for the Li10GeP2S12-type superionic conductors, The Journal of Chemical Physics 154(9) (2021).

[46] H. Wang, L. Zhang, J. Han, W. E, DeePMD-kit: A deep learning package for many-body potential energy representation and molecular dynamics, Computer Physics Communications 228 (2018) 178-184.

[47] S. Plimpton, Fast Parallel Algorithms for Short-Range Molecular Dynamics, Journal of Computational Physics 117(1) (1995) 1-19.